\documentstyle[12pt]{article}
\scrollmode

\newcommand{\bm}[1]{\mbox{\boldmath $#1$}}
\newcommand{\rd}{{\rm d}}
\newcommand{\be}{\begin{equation}}
\newcommand{\ee}{\end{equation}}
\newcommand{\ba}{\begin{eqnarray}}
\newcommand{\ea}{\end{eqnarray}}
\newcommand{\lb}[1]{\label{#1}}
\newcommand{\bb}[1]{\bibitem{#1}}
\begin{document}
\begin{titlepage}
\setcounter{page}{1}
\title{Wormhole cosmic strings}
\author{G\'erard Cl\'ement\thanks{E-mail:
 GECL@CCR.JUSSIEU.FR} \\
\small Laboratoire de Gravitation et Cosmologie Relativistes
 \\
\small Universit\'e Pierre et Marie Curie, CNRS/URA769 \\
\small Tour 22-12, Bo\^{\i}te 142,
 4 place Jussieu, 75252 Paris cedex 05, France}
\bigskip
\date{\small February 16, 1995}
\maketitle
\begin{abstract}
We construct regular multi-wormhole solutions to a gravitating $\sigma$
model in three space-time dimensions, and extend these solutions to
cylindrical traversable wormholes in four and five dimensions. We then
discuss the possibility of identifying wormhole mouths in pairs to give
rise to Wheeler wormholes. Such an identification is consistent with the
original field equations only in the absence of the $\sigma$-model
source, but with possible naked cosmic string sources. The resulting
Wheeler wormhole space-times are flat outside the sources and may be
asymptotically Minkowskian.
\end{abstract}
\end{titlepage}
\section{Introduction}
The intriguing possibility that we might live in a multiply-connected
universe and be able to travel to distant galaxies through traversable
wormholes has been popularized by the analysis of Morris and Thorne \cite
{18}. Traversable wormholes may occur as solutions to the Einstein field
equations with suitable sources violating the weak energy condition. While
most Lorentzian
wormhole solutions discussed in the literature are spherically
symmetric \cite{19}, this is an unnecessary limitation, as stressed by
Visser \cite {20}. For instance, if the weak energy condition is relaxed
there might occur cylindrical wormholes, which from afar would appear as
cosmic strings. In \cite{1}, generalizing previous work in (2 + 1)
dimensions \cite{21} (see also \cite{22}), we have shown that an infinite
cylinder of exotic matter with equal negative energy density and longitudinal
stresses ($\mu=\tau_z<0$) and vanishing azimuthal stress ($\tau_\varphi=0$)
generates a symmetrical wormhole space-time, with two axes at spacelike
infinity. The metric both `outside' and `inside' the matter cylinder is
the well-known conical cosmic string metric \cite{2}, with a deficit angle
(the same on both sides) which can be chosen at will, independently of the
values of the cylinder parameters.

In the present work we wish to investigate cylindrical wormholes in a more
fundamental, purely field-theoretical model, that of an $O(3)$ non-linear
$\sigma$-model field coupled repulsively to gravity. We have previously
shown \cite{3} that in three space-time dimensions this model admits static
multi-wormhole solutions with two points at spacelike infinity. These
solutions can be promoted in a straightforward fashion to four-dimensional
multi-`wormhole cosmic string' space-times with two axes at spacelike
infinity. However the would-be regular multi-wormhole solutions constructed
in \cite{3} actually have conical singularities, which give rise
to naked cosmic strings in the four-dimensional case. In this paper, we show
how these extra singularities may be removed to yield genuine regular
multi-wormhole solutions. The corresponding four-dimensional space-time
metrics are asymptotic to the conical cosmic string metric, with a deficit
angle which can be positive, zero or negative according to the value of the
$\sigma$-model fundamental length. From our
$\sigma$-model wormhole space-times, we derive multi-wormhole solutions to
sourceless five-dimensional general relativity, and discuss briefly
the structure of the four-dimensional metric and electromagnetic fields
which result from Kaluza-Klein dimensional reduction.
We also discuss the possibility of
identifying wormhole mouths in pairs to give rise to Wheeler wormholes in a
space-time with only one axis at spacelike infinity. We find that such an
identification is consistent with the original field equations only in the
absence of the $\sigma$-model source, but with possible naked cosmic
string sources. Finally,
we discuss the extension of our results to the case where the $\sigma$-model
field is minimally coupled to a Chern-Simons gauge field \cite{7}.

\setcounter{equation}{0}
\section{Three-dimensional $\sigma$-model wormholes}
The $O(3)$ nonlinear $\sigma$ model in three space-time dimensions is defined
by the action
\be \lb{1}
S = \int \rd ^3 x \, \sqrt{|g|} \, \frac{1}{2} \, [g^{\mu \nu} \, \partial_\mu
\bm{\phi} \cdot \partial_\nu \bm{\phi} + \lambda \, (\bm{\phi}^2 - \nu^2)] \, ,
\ee
where the Lagrange multiplier $\lambda$ constrains the isovector field
$\bm{\phi}$ to vary on the two-sphere $\bm{\phi}^2 = \nu^2$. As first
shown in \cite{4}, this model admits static multi-soliton solutions in
a flat background space-time. We showed \cite{5} that these solutions
are actually independent of the background metric, which may be curved,
and we derived \cite{5,3} the soliton solutions to the coupled
Einstein-$\sigma$ system
\be \lb{2}
S = \int \rd^3 x \, \sqrt{|g|} \, \frac{1}{2} \, [-\frac{1}{\kappa} \,
g^{\mu \nu} \, R_{\mu \nu} \pm g^{\mu \nu} \, \partial_\mu \bm{\phi} \cdot
\partial_\nu \bm{\phi} \pm \lambda \, (\bm{\phi}^2 - \nu^2)] \,
\ee
($\kappa = 8 \pi G$), where the field $\bm{\phi}$ may be coupled attractively
(upper sign) or repulsively (lower sign) to gravity (both signs are
possible in three-dimensional gravity \cite{11}; the lower sign
arises naturally in the case where the action (\ref{2}) is obtained by
dimensional reduction from a five-dimensional Kaluza-Klein theory \cite{8}).
These gravitating
$\sigma$-model solitons were independently constructed in \cite{6}.

We briefly recall the construction of Ref.\ \cite{3}. The stereographic map
\be \lb{3}
\phi_1 + i \, \phi_2 = \frac{2 \, \nu \, \psi}{1 + |\psi|^2} \, , \,\,\,
\phi_3 = \nu \, \frac{1 - |\psi|^2}{1 + |\psi|^2} \, ,
\ee
projects the sphere $\bm{\phi}^2 = \nu^2$ on the complex $\psi$ plane. The
field equations derived from the action (\ref{2}) may then be written
\ba \lb{4}
& & R_{\mu \nu} = \pm \, 2 \, \kappa \, \nu^2 \, F \, (\partial_\mu
\psi^\ast \, \partial_\nu \psi + \partial _\nu \psi^\ast \, \partial_\mu
\psi) \, , \nonumber \\
& & \frac{1}{\sqrt{|g|}} \, \partial_\mu(\sqrt{|g|} \, g^{\mu \nu} \,
\partial_\nu \psi) = 2 \, F^{1/2} \, \psi^\ast \, g^{\mu \nu} \,
\partial_\mu \psi \, \partial_\nu \psi \, ,
\ea
where $F(|\psi|) \equiv (1+|\psi|^2)^{-2}$.
We search for static solutions such that $\psi$ is time independent, and
the metric may be written
\be \lb{5}
\rd s^2 = h^2 \, \rd t^2 - {\rm e}^{2u} \, \rd \bm{x}^2
\ee
in isotropic spatial coordinates. The Einstein equations (\ref{4}) then
reduce to the system
\ba \lb{6}
& & \frac{\partial^2 h}{\partial \zeta \, \partial \zeta^\ast} = 0 \, ,
\nonumber \\
& & \frac{\partial^2 u}{\partial \zeta \, \partial \zeta^\ast} = \mp \,
\kappa \, \nu^2 \, F \, \left ( \left | \frac{\partial \psi}{\partial
\zeta} \right |^2 + \left | \frac{\partial \psi}{\partial \zeta^\ast}
\right |^2 \right ) \, , \nonumber \\
& & \frac{\partial}{\partial \zeta} \, \left ( {\rm e}^{-2u} \,
\frac{\partial h}{\partial \zeta} \right ) \pm 4 \, \kappa \, \nu^2 \,
h \, {\rm e}^{-2u} \, F \, \frac{\partial \psi^\ast}{\partial \zeta} \,
\frac{\partial \psi}{\partial \zeta} = 0 \, .
\ea
where $\zeta \equiv x + iy$ . In the case of multi-soliton solutions, the
metric should be asymptotic to that generated by a system of point particles.
This implies that the harmonic function $h$ is constant; we choose $h=1$. The
last equation (\ref{6}) then shows that $\psi$ must be an analytic or
anti-analytic function, which also solves the last equation (\ref{4})
(both sides vanish). We assume for definiteness $\psi$ to be analytic,
$\psi=\psi(\zeta)$, and, without loss of generality, we choose the South
pole of the sphere $\bm{\phi}^2=\nu^2$ (the centre of the stereographic
projection (\ref{3})) to be the image of the point at infinity of the
$\zeta$ plane, i.\ e.\ $\psi(\infty)=\infty$. Finally, the integration
of the second equation (\ref{6}) leads to the metric function
\be \lb{7}
{\rm e}^{2u} = \frac{(1 + |\psi|^2)^{\mp 2 \kappa \nu^2}}{|f(\psi)|^2} \, ,
\ee
where $f$ is an arbitrary analytical function of $\psi$.

If this function is constant then the metric
\be \lb{8}
\rd s^2 = \rd t^2 - (1 + |\psi|^2)^{\mp 2 \kappa \nu^2} \, \rd \zeta \, \rd
\zeta^\ast
\ee
is everywhere regular provided the point $\zeta = \infty$ is indeed at
spatial infinity. Assuming $\psi$ to be of the order of $\zeta^n$ ($n$
integer) for $\zeta \rightarrow \infty$, we find that for the upper sign
in (\ref{8}) the spatial metric is asymptotically conical (or cylindrical)
if $n \kappa \nu^2 \leq 1/2$, and compact if $n \kappa \nu^2 = 1$ (in the
case where $\psi$ is linear in $\zeta$, this is the well-known $\sigma$-model
monopole compactification mechanism \cite{9}), while for the lower sign the
metric is asymptotically pseudo-conical (the deficit angle is negative).

If on the other hand the function $f(\psi)$ is not constant, then it has at
least one zero $\psi_0$, which leads to a metric singularity unless the point
$\psi = \psi_0$ is at spatial infinity. There are then generically two points
at spatial infinity, $\psi = \infty$ and $\psi = \psi_0$. Following Ref.\
\cite{3}, we assume that these two points are respectively the South and
North pole of the sphere $\bm{\phi}^2 = \nu^2$ (implying $\psi_0 = 0$),
and that the North-South reflection
\be \lb{9}
\psi \rightarrow (\psi^\ast)^{-1}
\ee
about the equatorial plane $\phi_3 = 0$ is an isometry of our space-time.
Putting
\be \lb{10}
\psi = {\rm e}^Z \, ,
\ee
with $Z = X + iY$, this condition leads to the space-time metric in $Z$
coordinates
\be \lb{11}
\rd s^2 = \rd t^2 - (\cosh X)^{\mp 2 \kappa \nu^2} \, \frac{\rd Z  \, \rd
Z^\ast}{|g(Z)|^2} \, ,
\ee
which is invariant under (\ref{9}) if $|g(-Z)| = |g(Z)|$.

Now, because of the isometry (\ref{9}), we may take $\psi(\zeta)$ to be a
conformal map of the $\zeta$ plane on the South (or North) hemisphere,
i.\ e.\ the exterior (or the interior) of the circle $|\psi| = 1$ ($X=0$).
Such a map is given by the transformation \cite{12}
\be \lb{12}
\cosh Z \equiv \frac{1}{2} \, (\psi + \frac{1}{\psi}) = \zeta_1 \, ,
\ee
leading to the one-soliton metric,
\be \lb{13}
\rd s^2 = \rd t^2 - (\cosh X)^{\mp 2 \kappa \nu^2} \, |g(Z)|^{-2} \,
\frac{\rd \zeta_1 \, \rd \zeta_1^\ast}{|\zeta_1^2 - 1|} \, .
\ee
This is singular at the zeroes or poles of $g(Z)$, so that a necessary
condition for regularity is $g(Z)={\rm constant}$. However, even with
this choice, the metric (\ref{13}) still has two conical singularities
(branch points)
with angular deficit $\pi$ located at the two points $P\,(\zeta_1=-1)$ and
$P'\,(\zeta_1=1)$. In the conventional interpretation of conical singularities
in three-dimensional gravity \cite{10,11}, each of these singularities would
be associated with a point mass $m=\pi / \kappa$. Our point of view here is
that these singularities are spurious, and may be removed by transforming to
a suitable coordinate system,
 thereby revealing the wormhole structure of our space-time. The
 transformation (\ref{12}) maps the region $|\psi|>1$ of the $\psi$
 plane on the $\zeta_1$ plane cut along the segment $PP'$. The inverse
 transformation defines the bi-valued function $\psi(\zeta_1)$ which
 becomes single-valued on the Riemann surface made of two copies of the
 $\zeta_1$ plane connected along the cut $X=0$. The metric transformed
 from (\ref{13}) (with $g(Z)=l^{-1}$, $l$ constant) by $\psi(\zeta_1)$ is
\be \lb{17}
\rd s_1^2 = \rd t^2 - l^2 \, (1 + r^2)^{\mp \kappa \nu^2 - 1} \, [\rd r^2 +
(1 + r^2) \, \rd \theta^2] \, ,
\ee
where we have used the polar representation $\psi = R \, {\rm e}^{i \theta}$,
and $r = \frac{1}{2} (R - 1/R)$ varies from $-\infty$ to $+\infty$. In the
case of the upper sign, the points $r = \pm \infty$ are actually at a
finite distance, and the spatial sections are compact; they are
regular only for $\kappa \nu^2=1$, in which case we again recover the
$\sigma$-model spherical compactification. In the case of the lower sign,
the spatial sections have the two points at infinity $r = \pm \infty$,
the regular metric (\ref{17}) being asymptotically conical (for $\kappa
\nu^2<1$) or pseudo-conical (for $\kappa \nu^2>1$),
and locally cylindrical on the equator $r =0$.

In the limit $\kappa \nu^2 \rightarrow 0$ we recover the cylindrical
space-time $\rd s^2=\rd t^2-\rd r^2-\l^2\rd \theta^2$. Conversely,
the metric (\ref{13}) with $g=l^{-1}$,
$\kappa \nu^2=0$ is obtained
from the cylindrical metric by pinching the cylinder of radius $l$ along a
parallel. By cutting the cylinder along the resulting segment, we obtain two
copies of the bi-cone
 generated by two point particles $P$ and $P'$ of mass $\pi/\kappa$. The
 generic bi-cone is singular because it can be flattened only by making
two cuts extending from each particle to infinity; our construction shows
that the  bi-cone  can be maximally extended to a regular surface
---a cylinder--- when the two deficit angles are equal to $\pi$ (Fig.\ 1).

The multi-soliton solution is obtained from (\ref{12}) by the conformal map
\be \lb{14}
\zeta_1 = \prod_{i=1}^n \, (\zeta - a_i) \, ,
\ee
depending on $n$ complex constants $a_i$.
The associated metric (\ref{13}) is singular at the ($n-1$) zeroes of the
polynomial $\partial \zeta_1 / \partial \zeta$, unless the even function
$g(Z)$ is chosen precisely so as to compensate these zeroes, $g(Z) =l^{-1}
\partial \zeta_1 / \partial \zeta$ (in \cite{3}, $g(Z)$ was implicitely
assumed to be constant, so that the multi-soliton metric was actually
singular). The resulting multi-soliton, multi-wormhole solution is given
by equations (\ref{12}), (\ref{14}), and
\be \lb{15}
\rd s^2 = \rd t^2 - l^2 \, (\cosh X)^{2 \kappa \nu^2} \, \frac{\rd \zeta
\, \rd \zeta^\ast}{|\zeta_1^2 - 1|} \, ,
\ee
where we have taken the lower sign in (\ref{13}) (the upper sign leads to
singular solutions for all $n>1$). The metric (\ref{15}) contains $2n$
conical singularities located at the zeroes $b^{\pm}_i$ of $(\zeta^2_1-1)$,
each with angular deficit $\pi$. As in the case $n=1$, these metrical
singularities are characteristic of the wormhole topology, and may be
removed pairwise by transforming to a suitable coordinate system. The
spatial sections of the $n$-wormhole space-time are Riemann surfaces
made of two copies of the
 $\zeta_1$ plane joined along $n$ cuts, the $n$ components (assumed to
 be disjoint) of the equator $X=0$, each of which connects two
 singularities $\zeta=b^+_i$ ($\zeta_1=+1$) and $\zeta=b^-_i$
 ($\zeta_1=-1$). To remove any given pair of singularities
 ($b^-_i,b^+_i$), we make the coordinate transformation from
 $\zeta$ to $\tilde{\psi}$, defined by
\be \lb{18}
\zeta - \bar{b_i} = \frac{1}{2} \, (b^+_i - b^-_i) \, \tilde{\zeta_1}
= \frac{1}{4} \, (b^+_i - b^-_i) \, (\tilde{\psi} +
\frac{1}{\tilde{\psi}})
\ee
(where $\bar{b_i}=\frac{1}{2}(b^+_i + b^-_i)$); this maps the $i^{th}$
cut (which we now choose to be the straight segment connecting $b^+_i$
to $b^-_i$) into the circle $|\tilde{\psi}|=1$, in the vicinity of which
the transformed metric is regular and can be extended from
$|\tilde{\psi}|>1$ to $|\tilde{\psi}|<1$. The asymptotic behaviour
of the metric function in (\ref{15}) is
\be \lb{16}
{\rm e}^{2u} \sim \rho^{2n(\kappa \nu^2 - 1)} \,\,\,\,\,\,\,\,\,\,\,
\,\,\,\,\,\, (\zeta \rightarrow \infty)
\ee
$(\rho=|\zeta|)$. It follows that the spatial sections are, in each
Riemann sheet, asymptotically pseudo-conical for $\kappa \nu^2 >1$,
asymptotically Euclidean for $\kappa \nu^2 =1$, and asymptotically
conical (cylindrical) for $1-1/n \leq \kappa \nu^2 < 1$. The values
$\kappa \nu^2 = 1-2/n$ yield regular compact spatial sections of genus
$n-1$. For $\kappa \nu^2 = 0$, $n=2$, the maximally extended
spatial sections of (\ref{15}) are flat tori, as may be checked by
transforming the metric (\ref{15}) (where we can choose
$\zeta_1=\zeta^2-a^2$, with $a > 1$) to the Minkowski form
\be \lb{99}
\rd s^2 = \rd t^2 - \rd w \, \rd w^\ast
\ee
with
\be \lb{19}
\rd w = \frac{l \, \rd \zeta}{\sqrt{(\zeta^2-a^2)^2-1}} \, ,
\ee
and noting that the inverse function $\zeta (w)$ \cite{12} is a
bi-periodical Jacobi function,
\be \lb{98}
\zeta = \sqrt{a^2 + 1} \; {\rm sn} \left( \frac{\sqrt{a^2 + 1}}{l}
\, w, \, k \right)
\ee
with $k^2  = (a^2 - 1)/(a^2 + 1)$, which implies that the real and
imaginary part of $w$ should both be periodically
identified; conversely, the metric (\ref{15}) may be obtained
by pinching the flat torus $S^1 \times S^1$ along two opposite circles,
yielding two copies of the tetra-cone.

The total energy associated with a static metric of the form (\ref{5})
with $h=1$ may be determined from the asymptotic behaviour of the metric
function $u$ by \cite{5,11}
\be \lb{20}
M = - \frac{2 \, \pi}{\kappa} \, \lim_{\rho \rightarrow \infty} \, \rho
\, \frac{\partial u}{\partial \rho} \, .
\ee
In the case of a space-time with $n$ wormholes, this total energy is
related to the Euler invariant
\be \lb{21}
I = \frac{1}{16 \, \pi} \, \int_\Sigma \rd ^2 x \, \sqrt{|g|} \, g^{ij}
\, R_{ij}
\ee
(where the integral extends over both sheets of the Riemann surface
$\Sigma$) by\footnote{Eq.\ (\ref{22}), obtained by using the Gauss-Bonnet
theorem (see also \cite{6}), is valid for the case where
$\Sigma$ has two asymptotically flat regions with the same angular
deficit. In the case of only one asymptotically flat region, the
left-hand side of (\ref{22}) should be replaced by M.} \cite{3}
\be \lb{22}
2 \, M = \frac{4 \, \pi}{\kappa} \, (n-2 \, I) \, .
\ee
For our $\sigma$-model regular wormhole metrics (\ref{15}), the Euler
invariant $I$ is, from the first equation (\ref{4}), proportional to
the soliton number \cite{5} (degree of the map $\bar{\Sigma} \rightarrow
S^2$, where $\bar{\Sigma}$ is the closure of $\Sigma$), equal to the
wormhole number,
\be \lb{23}
I = \frac{\kappa \, \nu^2}{2} \, \frac{1}{\pi} \, \int_\Sigma \rd^2 x
\, F \, \left |\frac{\partial \psi}{\partial \zeta} \right |^2 = n \,
\frac{\kappa \, \nu^2}{2} \, .
\ee
The total energy given by eq.\ (\ref{22}) is therefore the energy of a
system of $n$ non-interacting particles,
\be \lb{24}
M = n \, (1 - \kappa \, \nu^2) \, \frac{2 \, \pi}{\kappa} \, ,
\ee
in accordance with the asymptotic behaviour (\ref{16}).

\setcounter{equation}{0}
\section{Wormhole cosmic strings in four and five dimensions}
Our three-dimensional multi-wormhole space-times of metric (\ref{15})
can be factored
by the $z$-axis, leading to four-dimensional multi-wormhole-cosmic
string space-times of metric
\be \lb{25}
\rd s^2 = \rd t^2 - l^2 \, (\cosh X)^{2 \kappa \nu^2} \,
\frac{\rd \zeta \, \rd \zeta^\ast}{|\zeta_1^2 - 1|} - \rd z^2 \, .
\ee
For an observer at spacelike infinity, this appears to be a static
system of $n$ parallel cosmic strings with mass per unit length and
longitudinal tension both equal to $(1-\kappa \nu^2) 2 \pi / \kappa$.
However at short range each individual `cosmic string' turns out to be
a cylindrical wormhole leading to the other sheet of three-dimensional
 space.

The regular metric (\ref{25}) may also be generalized to a hybrid
 system of $n$ wormhole cosmic strings together with $p$ naked cosmic
 strings (line singularities) in each sheet of three-dimensional space
 by choosing appropriately the function $g(Z)$ in eq.\ (\ref{13}).
 These systems survive in the limit of a vanishing $\sigma$-model
 source ($\kappa \nu^2 \rightarrow 0$), the naked cosmic strings acting
 as sources for the multi-wormhole configuration; the total energy per unit
 length (tension) is then
\be \lb{26}
M = \frac{2 \, n \, \pi}{\kappa} + \sum_{i=1}^p m_i
\ee
(where $m_i$ is the mass per unit length of the $i^{th}$ naked cosmic
string), so that a necessary condition for three-dimensional space to
be open ($M \leq 2 \pi / \kappa$) is $\sum m_i \leq 0$.

For the special value $\kappa \nu^2 =2$, Kaluza-Klein cosmic string
space-times with negative deficit angle may also be derived from our
$\sigma$-model wormhole space-times. Making for five-dimensional
general relativity with three commuting Killing vectors the ansatz
\cite{8,3}
\be \lb{27}
\rd s^2 = g_{ij}(x^k) \, \rd x^i \, \rd x^j + (2 \, \phi_a (x^k) \,
\phi_b (x^k) - \delta _{ab}) \, \rd x^a \, \rd x^b
\ee
($i,j,k=1,2;\,\, a,b=3,4,5$) where $\bm{\phi}$ varies on the unit
sphere ($\bm{\phi}^2=1$), we reduce the five-dimensional
Einstein-Hilbert action to
\be \lb{28}
S = \int \rd^3 x \, \int \rd^2 x \, \sqrt{|g|} \, \frac{1}{2}
\, g^{ij} \, [- \frac{1}{\kappa} \, R_{ij} - \frac{2}{\kappa}
\, \partial _i \bm{\phi} \cdot \partial _j \bm{\phi}] \, .
\ee
After a suitable rescaling of $\bm{\phi}$, this is equivalent
to the static restriction of the action (\ref{2}) where the lower
sign is taken, and $\kappa \nu^2 = 2$. Accordingly, to each
multi-wormhole solution (\ref{14}), (\ref{15}), there corresponds
a solution of sourceless five-dimensional general relativity. The
resulting metric (\ref{27}) has the signature $- - - + -$ at spacelike
infinity if the stereographic map is chosen to be for instance
\be \lb{29}
\phi_5 + i \phi_3 = \frac{2 \, \psi}{1+|\psi|^2} \, , \,\,\,\, \phi_4
= \frac{1-|\psi|^2}{1+|\psi|^2} \, .
\ee
When $\psi$ varies from $\psi=\infty$ to $\psi=0$ (the points at
infinity of the two Riemann sheets of $\Sigma$), $\phi_4$ varies
from $-1$ to $+1$, so that light cones tumble over from future-oriented
for e.\ g.\ $\psi=\infty$, to space-oriented (in the plane $x^3,x^5$)
on the cuts $|\psi|=1$, and to past-oriented for $\psi=0$. This shows
that our Kaluza-Klein wormhole space-times are metrical kinks \cite{14},
with kink number $n$.

The Kaluza-Klein projection of the five-dimensional metric (\ref{27})
leads to the four-dimensional metric components, the electromagnetic
potentials and the scalar field
\be \lb{30}
\bar{g}_{ab} = \frac{2 \, \phi_a \, \phi_b}{1 - 2 \, \phi_5^2} -
\delta_{ab} \, , \,\,\, A_a = \frac{2 \, \phi_a \, \phi_5}{1 - 2
\, \phi_5^2} \, , \,\,\, \sigma = 1 - 2 \, \phi_5^2 \, .
\ee
The $n=1$ fields are not axisymmetric. The electric potential $A_4$
and the metric tensor component $\bar{g}_{34}$ are asymptotically
`cylindrical dipole' fields (gradients of the two-dimensional
monopole harmonic field $\ln \rho$), while the other fields are
asymptotically `cylindrical quadrupole'. While the five-dimensional
metric (\ref{27}) is everywhere regular, the Kaluza-Klein projection
procedure breaks down for $\phi_5^2 = 1/2$, i.\ e.\ in the case
$n = 1$ on the two oval cylinders (which may be thought of as the two
effective `sources' for the four-dimensional fields (\ref{30}))
$\cos \theta = \pm \sqrt{2} \, (1+\rho^2)/3\rho$
(where $\zeta = \rho {\rm e}^{i \theta}$), inside which the
five-dimensional geometry with
compactified fifth dimension admits the closed timelike curves $x^\mu
= {\rm constant}$.

\newpage
\setcounter{equation}{0}
\section{Wheeler wormholes}
In the second section we have explained how the spatial part of the metric
(\ref{15}) may be maximally extended to a manifold with two asymptotically
flat regions connected by $n$ traversable wormholes. Wheeler wormholes
\cite{17}, by contrast, connect two distant regions of a spatial manifold
with only one asymptotically flat region (Fig.\ 2). It is often taken for
granted that two-sided wormhole systems may be transformed to Wheeler
wormhole systems by suitably identifying together the two asymptotically
flat regions, although it is far from obvious that this can be done
consistently. Let us discuss how such an identification, which gives rise
to a new maximal extension of the metric (\ref{15}), may be carried out in
our model for the case $n = 2$.

For $n = 2$, we may always choose a coordinate system such that the map
(\ref{14}) simplifies to
\be \label{35}
\zeta_1 = \zeta^2 - a^2
\ee
($a > 1$). With this parametrization the metric (\ref{15}) is manifestly
invariant under the symmetry $ \zeta \rightarrow -\zeta$, which exchanges
the two cuts. Combining this isometry with the complex inversion $\psi
\rightarrow 1/\psi$, which exchanges the two Riemann sheets, we obtain an
isometry which maps a neighbourhood of the left-hand cut in the second
Riemann sheet into a neighbourhood of the right-hand cut in the first
Riemann sheet, and vice-versa. This means that we can do away with the
second Riemann sheet altogether, so that the spatial sections are obtained
from the first Riemann sheet alone by identifying the two cuts. We thus
arrive at the reinterpretation of our wormhole pair as a single Wheeler
wormhole with only one point at spatial infinity.

This interpretation may be checked out at the geodesic level in the
special case $\kappa \nu^2 = 0$. We have shown that in this case the
two-sheeted extension of the metric (\ref{15}) leads to the toroidal
space-time (\ref{99}), with geodesics $w - w_0 = \beta t$. A large-circle
geodesic ${\cal I}m \, w = {\rm constant}$ crosses the two cuts (two
opposite small circles of the torus), going for instance from the
right-hand cut to the left-hand cut in the upper half of the first Riemann
sheet, then back to the right-hand cut in the lower half of the second
Riemann sheet (Fig.\ 3). It follows from the identity ${\rm sn}(u + 2K) =
- {\rm sn}\,u$, where $4K$ is the real period of the Jacobi function in
eq. (\ref{98}), that the transformation $\zeta \rightarrow -\zeta$, $\psi
\rightarrow 1/\psi$ maps each point $P'$ of the second half of the
geodesic into the symmetrical point $P$ on the first half of the geodesic,
i.\ e.\ (Fig. 4) maps the torus pinched along two symmetrical circles into
a smaller torus pinched along a single circle (the tetra-cone viewed as
its own maximal extension). The topological picture is the same in the
non-compact case ($\kappa \nu^2 \ge 1/2$), the two-sheeted spatial
sections being compactified by adding two symmetrical points at infinity,
which are identified together in the Wheeler wormhole interpretation.

However there is a price to pay for this reinterpretation. The complex
inversion $\psi \rightarrow 1/\psi$ corresponds, from eq. (\ref{3}), to
the transformation of the spherical scalar field
\be \lb{36}
(\phi_1, \,\phi_2, \,\phi_3) \rightarrow (\phi_1, \,-\phi_2, \,-\phi_3),
\ee
so that our isometrical identification between the two Riemann sheets
would lead to an identification between two inequivalent matter field
configurations. In other words, the identification we have just described
is possible geometrically, but not ---in the case of a $\sigma$-model
source--- physically. Neither is this identification possible for the
purely geometrical five-dimensional model of Sect. 3 ($\kappa \nu^2 = 2$),
as it does not lead to an isometry for the five-dimensional geometry
(\ref{27}). Indeed, on account of eqs. (\ref{29}) and (\ref{30}) the
transformation $\psi \rightarrow 1/\psi$ reverses the electromagnetic
potentials, i.\ e.\ is equivalent to charge conjugation, so that our
hypothetical Kaluza-Klein Wheeler wormhole would not conserve electric
charge.

We conclude that our construction of Wheeler wormholes is feasible only in
the absence of the $\sigma$-model source considered in this paper, $\kappa
\nu^2 = 0$. For instance, the four-dimensional metric
\be \lb{37}
{\rm d}s^2 = {\rm d}t^2 \,- \,|g(\zeta)|^2 \frac{{\rm d}\zeta \,
{\rm d}\zeta ^\ast}{|(\zeta^2-a^2)^2 \,- \,1|} \,- \,{\rm d}z^2
\ee
(where $g(\zeta)$ has $p$ zeroes, and $|g(-\zeta)| = |g(\zeta)|)$ can thus
be maximally extended to a Wheeler wormhole cosmic string generated by p
naked cosmic strings of negative masses per unit length (or tensions) $m_i$.
The net mass per unit length of this Wheeler wormhole (defined from the
asymptotic deficit angle) is
\be \lb{38}
M \,= \, \frac{4 \pi}{\kappa} \,+ \,\sum_{i=1}^p m_i\, .
\ee
The simplest case $p=1$, $g(\zeta) = {\rm const.} \, \zeta^{- \kappa m/2
\pi}$ corresponds to a single naked straight string generating a Wheeler
wormhole with a net mass which may be positive provided $-4\pi /
\kappa < m \le -2\pi / \kappa$, and zero for $m = -4 \pi / \kappa$
(the flat metric (\ref{37}) is then asymptotically Minkowskian).

\setcounter{equation}{0}
\section{Discussion}
We have constructed regular multi-wormhole solutions to an antigravitating
$\sigma$ model in three space-time dimensions, and extended these
solutions to cylindrical wormholes in four and five dimensions. We have
also discussed how a pair of two-sided wormholes may be reinterpreted as a
Wheeler wormhole. However this reinterpretation is consistent with the field
equations only in the absence of the $\sigma$-model source (three or four
dimensions).

While our emphasis in this paper was on regular solutions, an interesting
by-product of our analysis is the construction of cylindrical wormholes in
four dimensions with naked cosmic string sources. Visser \cite{20}
previously suggested the existence of flat-space wormholes framed by
cosmic string configurations. Our construction goes beyond Visser's in
two respects. First, our flat-space wormholes are not framed by the naked
cosmic strings, which can be far from the wormhole mouths if $a \gg 1$.
Second, we are able to construct not only two-sided wormholes, but also a
Wheeler wormhole (with only one asymptotically Minkowskian region)
generated by a single naked straight string.

Finally, let us mention that our construction of multi-wormhole solutions
may be straightforwardly extended to the case of a $\sigma$ model gauged
with a Chern-Simons gauge field \cite{7}. This model is defined by the
action, which replaces (\ref{1}),
\ba \lb{31}
S & = & \int \rd ^3 x \, \sqrt{|g|} \, \frac{1}{2} \, [g^{\mu \nu} \,
D_\mu \bm{\phi} \cdot D_\nu \bm{\phi} + \lambda \, (\bm{\phi}^2 - \nu^2)
\nonumber \\
& & \mbox{} - \mu \, \frac{1}{\sqrt{|g|}} \, \varepsilon^{\mu \nu \rho}
\, (\partial_\mu \bm{A}_\nu \cdot \bm{A}_\rho + \frac{1}{3} \,
\varepsilon^{abc} \, A_\mu^a \, A_\nu^b \, A_\rho^c)] \, ,
\ea
where $\varepsilon^{\mu \nu \rho}$ is the antisymmetric symbol, and
$D_\mu \phi^a = \partial _\mu \phi^a + \varepsilon ^{abc} A_\mu^b \phi^c$
the gauge covariant derivative. As shown in \cite{7} for the case of a
flat background space-time, the static finite-energy solutions of the
model (\ref{31}) are given, up to a gauge transformation, by
\be \lb{32}
\bm{A}_i = 0 \, , \,\,\,\, \bm{A}_0 = \eta \, \frac{1}{\mu} \, \bm{\phi} \, ,
\ee
where $\bm{\phi}(\bm{x})$ is a static finite-energy solution of the model
(\ref{1}), and $\eta = \pm 1$. This result extends trivially to the case
of the gravitating gauged $\sigma$ model
\ba \lb{33}
S & = & \int \rd ^3 x \, \frac{1}{2} \, \{ \sqrt{|g|} \,
[- \frac{1}{\kappa} \, g^{\mu \nu} \, R_{\mu \nu} \pm g^{\mu \nu}
\, D_\mu \bm{\phi} \cdot D_\nu \bm{\phi} \nonumber \\
& & \mbox{} \pm \lambda \, (\bm{\phi}^2 - \nu^2)] \mp \mu \,
\varepsilon^{\mu \nu \rho} \, (\partial_\mu \bm{A}_\nu \cdot
\bm{A}_\rho + \frac{1}{3} \, \varepsilon^{abc} \, A_\mu^a \,
A_\nu^b \, A_\rho^c)] \} \, ,
\ea
because the Chern-Simons term in (\ref{33}) is not coupled to gravity,
while from eq.\ (\ref{32}) $D_\mu \bm{\phi} = \partial _\mu \bm{\phi}$,
so that the gauged field equations for the gravitational and scalar fields
reduce to the ungauged equations (\ref{4}). It follows that (in the case
of the lower sign in (\ref{33})) our multi-soliton multi-wormhole solutions
(\ref{14}), (\ref{15}) yield multi-magnetic vortex configurations \cite{7}
for the effective Abelian electromagnetic field
\be \lb{34}
{\cal F}_{\mu \nu} = \bm{\phi} \cdot \bm{F}_{\mu \nu} - \varepsilon ^{abc}
\, D_\mu \phi^a \, D_\nu \phi^b \, \phi^c
\ee
(a four-dimensional equivalent is the Einstein-Yang-Mills-Higgs
wormhole-monopole constructed in \cite{16}).

\newpage
\noindent{\Large \bf Figure captions}
\begin{description}
\item[Fig.\ 1:]
Two possible definitions of the domain of analyticity of the metric
(\ref{13}). For $\kappa \nu^2 = 0$ the first possibility leads to a
bi-cone, while the second possibility leads, after analytical continuation,
to a cylinder.
\item[Fig.\ 2:]
A Wheeler wormhole.
\item[Fig.\ 3:]
An ${\cal I}m \,w = {\rm const.}$ large circle of the torus crosses the
two cuts. The full portion of the geodesic is in the first Riemann sheet,
the dashed portion in the second Riemann sheet. The identification of the
symmetrical points $P'$ and $P$ results in a Wheeler wormhole.
\item[Fig.\ 4:]
The isometrical identification of the two halves of the torus results in a
smaller torus with only one pinch.
\end{description}
\end{document}